\documentclass[aps,prb,nobibnotes,citeautoscript,superscriptaddress,footinbib,twocolumn,10pt,balancelastpage]{revtex4-1}
\usepackage[latin1]{inputenc}
\usepackage[T1]{fontenc}
\usepackage{amsmath,amssymb}
\usepackage{mathrsfs} 
\usepackage{graphicx,color,colortbl}
\usepackage{rotating,array,tabularx,booktabs}
\usepackage{varwidth,xcolor}
\usepackage{placeins}
\usepackage{siunitx}
\usepackage{ifthen}
\usepackage[normalem]{ulem}
\usepackage{multirow}
\DeclareGraphicsExtensions{.pdf,.PDF,.png,.PNG}

\allowdisplaybreaks

\let\originalcite\cite
\newboolean{WITHINREMOVE}\setboolean{WITHINREMOVE}{false}%
\renewcommand{\cite}[1]{\ifthenelse{\boolean{WITHINREMOVE}}{\mbox{\originalcite{#1}}}{\originalcite{#1}}}%

\begin{document}

\title{Acoustic neural networks: Identifying design principles and exploring physical feasibility}
\author{Ivan Kalthoff}
\affiliation{Department of Physics, RWTH Aachen University, 52074 Aachen, Germany}
\affiliation{DWI -- Leibniz Institute for Interactive Materials, 52074 Aachen, Germany}
\affiliation{Institute of Theoretical Physics, Center for Soft Nanoscience, University of M{\"u}nster, 48149 M{\"u}nster, Germany}

\author{Marcel Rey}
\affiliation{Institute of Physical Chemistry, Center for Soft Nanoscience, University of M{\"u}nster, 48149 M{\"u}nster, Germany}

\author{Raphael Wittkowski}
\email[Corresponding author: ]{rgwitt25@dwi.rwth-aachen.de}
\affiliation{Department of Physics, RWTH Aachen University, 52074 Aachen, Germany}
\affiliation{DWI -- Leibniz Institute for Interactive Materials, 52074 Aachen, Germany}
\affiliation{Institute of Theoretical Physics, Center for Soft Nanoscience, University of M{\"u}nster, 48149 M{\"u}nster, Germany}

\begin{abstract}
Wave-guide-based physical systems provide a promising route toward energy-efficient analog computing beyond traditional electronics. Within this landscape, acoustic neural networks represent a  promising approach for achieving low-power computation in environments where electronics are inefficient or limited, yet their systematic design has remained largely unexplored. Here we introduce a framework for designing and simulating acoustic neural networks, which perform computation through the propagation of sound waves. Using a digital-twin approach, we train conventional neural network architectures under physically motivated constraints including non-negative signals and weights, the absence of bias terms, and nonlinearities compatible with intensity-based, non-negative acoustic signals. Our work provides a general framework for acoustic neural networks that connects learnable network components directly to physically measurable acoustic properties, enabling the systematic design of realizable acoustic computing systems. We demonstrate that constrained recurrent and hierarchical architectures can perform accurate speech classification, and we propose the SincHSRNN, a hybrid model that combines learnable acoustic bandpass filters with hierarchical temporal processing. The SincHSRNN achieves up to 95\% accuracy on the AudioMNIST dataset while remaining compatible with passive acoustic components. Beyond computational performance, the learned parameters correspond to measurable material and geometric properties such as attenuation and transmission. Our results establish general design principles for physically realizable acoustic neural networks and outline a pathway toward low-power, wave-based neural computing.
\end{abstract}
\maketitle

\section{Introduction}
Deep learning has transformed fields such as speech recognition, image classification, and scientific data analysis \cite{LeCun2015}. Its rapid progress has been enabled primarily by advances in hardware rather than fundamentally new algorithms \cite{Thompson2020}. However, sustaining this trajectory is increasingly difficult as transistor miniaturization reaches physical limits \cite{Waldrop2016}. This has motivated research into alternative paradigms such as neuromorphic, quantum, and wave-based computing \cite{Thompson2020,Markovic2020}.

In this context, \emph{acoustic neural networks}, architectures that process information via the propagation of sound waves rather than electronic signals, represent an emerging yet largely unexplored direction. Prior work in acoustic metamaterials has demonstrated that engineered wave structures can perform analog mathematical operations such as differentiation, equation solving, and edge detection, and even enable neural-like inference through passive meta-neural and mechanical neural networks\cite{Weng2020,Jiang2023,Lee2022,Silva2014, Zuo2018,Zangeneh2021}.
Together, these results demonstrate that wave-based physical systems can implement fundamental operations similar to those used in neural networks. In acoustic neural networks, the wave amplitude or intensity carries the signal, while material and geometric properties determine how signals combine and attenuate. These amplitude modulations act as the physical counterpart of neural network weights, setting the relative influence of different inputs. Similar to optical neural networks \cite{Lin2018,Xu2021}, acoustic neural networks can perform computations directly on wave signals without analog-to-digital conversion, reducing power consumption and latency \cite{Weng2020,Hughes2019,Zangeneh2021,Zolfagharinejad2025}. A key challenge remains the realization of nonlinear activation functions, which are essential for neural computation because they allow networks to represent complex, nonlinear relationships that cannot be captured by linear wave propagation alone. 
While nonlinear effects can also occur in conventional materials at high acoustic intensities, when the regime of linear elasticity is exceeded, their magnitude and tunability are generally limited. Recent advances in nonlinear acoustic materials and metamaterials, however, demonstrate stronger and design-controlled nonlinear effects, specifically in the form of amplitude-dependent attenuation, where the acoustic loss can increase, decrease, or saturate with growing intensity \cite{Fang2025,Fiore2020,Mork2022,Huang2020}, providing a promising physical mechanism for such nonlinear transformations.

Acoustic neural networks could open the door to low-power applications ranging from direct analog speech recognition and smart hearing aids to computing in environments where conventional electronics are limited \cite{Xiao2024,Li2024}. In the context of hearing aids, such systems could help identify relevant acoustic patterns, such as warning signals, while operating with minimal energy consumption. The demand for low-power, edge-based processing of acoustic information further motivates the design of neural architectures composed of acoustic elements \cite{Zolfagharinejad2025}.

Wave-based physical neural networks, including optical systems, acoustic and mechanical metamaterials, and analog recurrent structures, have demonstrated that wave propagation can perform learned transformations in hardware. However, within the acoustic domain in particular, existing implementations remain fixed-function or limited-operation systems, typically realizing a single trained mapping or isolated operations such as filtering, differentiation, or inference \cite{Weng2020,Jiang2023,Lee2022, Silva2014}. Despite partial reconfigurability or in-situ training in some systems, there is still no general framework for constructing full neural architectures whose parameters map directly onto physically realizable acoustic processes. Here, we address this gap by introducing a unified digital-twin approach that constrains all computations to passive, non-negative acoustic operations and whose parameters map directly onto measurable transmission, attenuation, and nonlinear loss. This perspective treats acoustic neural networks not as single engineered devices but as broadly applicable computational architectures grounded in the physics of sound. In turn, this framework enables the systematic design and optimization of future low-power acoustic neural hardware directly informed by the learned structure of the digital twin.

To realize this framework in practice, we develop constrained digital models that serve as blueprints for analog acoustic neural networks. Our models restrict signals and weights to non-negative values (representing sound intensity and attenuation), omit bias terms, and employ activation functions compatible with non-negative signals.
The inherently temporal nature of acoustic signals, together with the cumulative attenuation that arises during wave propagation, makes recurrent neural networks (RNNs) the simplest computational architecture that mirrors the dynamical behavior of physical acoustic systems. More expressive gated models such as Long Short-Term Memory (LSTM) or Gated Recurrent Unit (GRU) networks \cite{Hochreiter1997,Cho2014} rely on multiplicative gating mechanisms and signed activations that are incompatible with passive, non-negative acoustic media, making simple RNNs the physically realizable choice. Likewise, we evaluate all architectures directly on raw audio intensities, which correspond to the physical input available in an acoustic system and avoid digital preprocessing steps such as spectrogram computation that would break the analog-computing paradigm.

Building on this conceptual framework, we introduce a set of constrained architectures for speech classification on the AudioMNIST dataset \cite{Becker2024}. Starting from RNNs that capture temporal structure, we introduce hierarchical subsampling to handle long sequences, and finally propose a Hierarchical Subsampling Recurrent Neural Network with learnable sinc-based bandpass filters at the input stage (referred to as SincHSRNN). These sinc filters provide an interpretable and physically realizable mechanism for frequency-selective preprocessing, while the hierarchical recurrent component captures temporal dependencies. The SincHSRNN achieves competitive performance while remaining compatible with acoustic constraints.

The article is organized as follows: Sec.\ \ref{methods} introduces the underlying concept of acoustic neural networks, explains how neural operations can be mapped onto physically realizable acoustic components, and describes the digital-twin framework, network architectures, dataset and training procedures used to implement and evaluate the constrained models.
In Sec.\ \ref{results}, we present and discuss the results for recurrent, hierarchical, and sinc-based architectures, highlighting the influence of acoustic constraints on model performance and stability.
Finally, Sec.\ \ref{conclusions} summarizes the key findings and outlines design principles for future physically realizable acoustic neural networks.

\section{\label{methods}Methods}
\subsection{Mapping acoustic neural networks onto physical systems}
Acoustic neural networks aim to perform neural network computation through the physical propagation of sound waves, analogous to optical neural networks that use light for information processing \cite{Xu2021,Lin2018,Wang2022}. In such systems, the amplitude or intensity of an acoustic wave represents the signal transmitted between neurons, while the physical properties of the propagation medium determine the connection strength. This approach offers a path toward low-power, analog computing architectures that directly operate on acoustic signals with minimal digital processing.

\subsubsection{Mapping neural operations to acoustic processes}
\begin{figure}[htb]
\centering
\includegraphics[width=1\linewidth]{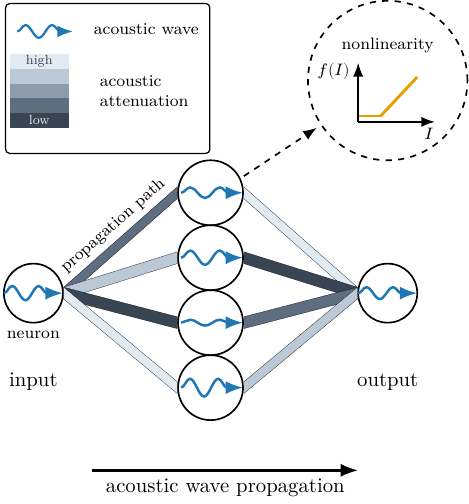}%
\caption{\label{fig:acoustic-concept} Conceptual schematic of an acoustic neural network. Information is transmitted and processed using sound waves instead of electrical signals. Propagation paths act as attenuating connections with transmission coefficients corresponding to neural weights ($w \in [0,1]$). Nonlinear transformations arise from intensity-dependent attenuation within acoustic media, serving as the activation function. Summation of signals occurs through the superposition of converging waves.}
\end{figure}

The conceptual structure of an acoustic neural network is illustrated in Fig.\ \ref{fig:acoustic-concept}. The functional elements of conventional neural networks can be mapped to acoustic processes. A connection weight corresponds to the transmission coefficient of an acoustic path, which depends on material attenuation, geometry, and reflection at interfaces. As acoustic systems are inherently passive, no amplification occurs, constraining weights to the range $w \in [0,1]$. The system considered here includes no external power source, and summation of incoming signals arises naturally from the superposition of converging waves. A bias term could, in principle, be implemented as a constant external sound source, but its realization would require an additional emitter per neuron and is therefore omitted for practical simplicity. 

Nonlinear activation functions can be introduced by exploiting intensity-dependent attenuation effects in nonlinear acoustic media. These mechanisms produce output amplitudes that vary nonlinearly with input intensity, analogous to activation functions in digital networks. In this way, a material with intensity-dependent transmission effectively acts as an acoustic neuron. To illustrate the plausibility of such nonlinear behavior in acoustic systems, an offset version of the rectified linear unit (ReLU)
\begin{equation}
f(I) = \max(I - I_{\mathrm{c}}, 0)
\end{equation}
can serve as an example of an activation suitable for non-negative neural networks \cite{Becker2023}, where $I$ denotes the input intensity and $I_{\mathrm{c}}$ a characteristic threshold. This function captures the behavior of an acoustic element whose transmission remains negligible for intensities below $I_{\mathrm{c}}$ and increases linearly once the threshold is exceeded. Such amplitude-gated transmission has been observed in nonlinear acoustic materials and metamaterials \cite{Mork2022,Huang2020}.

\subsubsection{Digital twin framework}
To explore this concept systematically, the networks studied in this work are designed as digital twins of physically realizable acoustic systems. Each digital parameter corresponds directly to an acoustic property: non-negative input signals represent sound intensity, weights are restricted to $[0,1]$ to model transmission losses, and nonlinearities are limited to functions compatible with non-negative signals realizable in known acoustic materials. Bias terms are excluded, and summation is implemented through linear superposition. It should be noted that the internal signal processing within the network is designed to be acoustically realizable. The output of the network, however, can be handled in different ways: one option is to process the final neuron signals digitally, for example by applying a softmax function to obtain class probabilities, while another is to use simple electronic components, such as comparators or multiplexers, to detect the strongest output signal and trigger a corresponding response. In both cases, the computationally demanding signal processing remains acoustic, whereas only the output stage involves minimal electronic operation.

The digital twin thus provides a simulation environment for identifying suitable architectures and parameter regimes. After training, the resulting network weights and structure define a blueprint for a potential acoustic implementation, guiding the selection of material properties, geometries, and nonlinear elements.

\subsection{Network architectures}
In choosing the model architectures, we restrict ourselves to recurrent structures whose computations can be mapped onto passive acoustic processes. Simple RNNs are compatible with these constraints because their recurrence relation requires only linear superposition, attenuation, and a scalar nonlinearity,  all operations that can be realized through acoustic transmission and intensity-dependent attenuation. Likewise, we design our models to operate directly on the non-negative acoustic intensity values that would be available in a physical system, rather than on digitally preprocessed representations such as spectrograms or  Mel-frequency cepstral coefficients (MFCCs). Avoiding such preprocessing ensures that all computations within the model correspond to transformations that could, in principle, be realized acoustically.
Three classes of neural network architectures were considered, each reflecting an increasing level of sophistication. In this work, model capacity refers to the number of trainable parameters determined by the network architecture, which increases with the number of hidden units and layers.

\subsubsection{Recurrent neural networks (RNNs)}
As a baseline for sequential modeling, simple RNNs were trained on the binary subset of AudioMNIST (digits 0 and 1). In constrained variants, weights were clamped to the range $[0,1]$ to model acoustic attenuation, bias terms were removed, and activation functions suitable for non-negative signals were employed, such as an offset version of the absolute value or ReLU function \cite{Becker2023}. These restrictions emulate the physical constraints of real acoustic media but limit model expressiveness and stability for long input sequences.
Each RNN consists of a single recurrent layer with 8 to 64 hidden units, followed by a fully connected output layer that maps the final hidden state to the class logits for the softmax classifier. The unconstrained model uses a standard tanh activation, whereas the constrained model employs the offset non-negative activations mentioned above. All weights are initialized from a uniform distribution, and hidden states are initialized with small random values in the same range. Detailed training parameters and hyperparameter settings for the RNN models are summarized in Appendix \ref{appendix}.

\subsubsection{Hierarchical subsampling recurrent neural networks (HSRNNs)}
To address the difficulty of modeling long raw audio sequences, HSRNNs \cite{Graves2012} were employed. These architectures reduce the temporal resolution between recurrent layers through hierarchical subsampling, effectively shortening sequence length while preserving relevant information. Each HSRNN consists of multiple recurrent layers with increasing hidden dimensions and a subsampling step between layers. For each layer, the input sequence is divided into non-overlapping segments defined by a predefined subsampling factor. Each element within a segment is processed by its own position-specific weight matrix, and the resulting outputs are summed and combined with the recurrent input from the previous hidden state. The unconstrained model uses the standard tanh activation, while the constrained model replaces it with an offset absolute or offset ReLU activation to maintain non-negative hidden states. The output sequence of one layer serves as the input to the next. After the final recurrent layer, a fully connected layer with tanh activation precedes a linear classifier that maps to the class logits for the softmax. Bias terms are disabled in the constrained model, and weights are initialized using Xavier-uniform initialization, with constrained variants enforcing non-negative weights through initialization and clamping. The complete set of training parameters and subsampling configurations used for the HSRNNs is provided in Appendix \ref{appendix}.

\subsubsection{Hierarchical subsampling recurrent neural networks with learnable sinc filters (SincHSRNNs)}

\begin{figure*}[htb]
\centering
\includegraphics[width=1\linewidth]{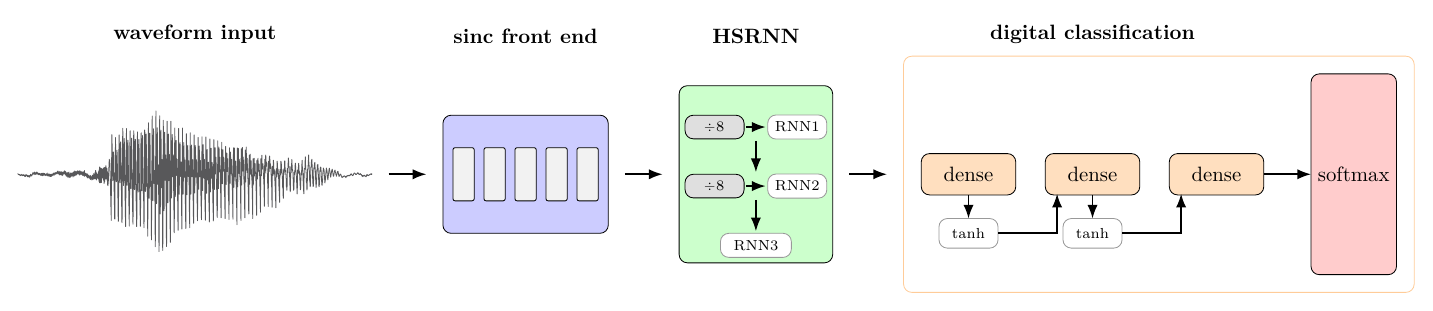}%
\caption{\label{fig:sincHSRNN-concept} Raw waveform input is processed by a front end of five learnable sinc filters that act as interpretable bandpass elements. The filtered signals are passed to a HSRNN composed of three to four recurrent layers, labeled RNN1--RNN3 in the figure, each preceded by temporal downsampling by a factor of 8 with learnable weights. The resulting feature representation is fed to two fully connected layers, denoted Dense in the figure, with $\tanh$ activations and a final fully connected output layer followed by a softmax classifier that produces the digit-class probabilities.}
\end{figure*}

Our main contribution is the SincHSRNN, which combines learnable sinc filters \cite{Ravanelli2018} at the input stage with the hierarchical recurrent structure. The overall architecture of the proposed model is illustrated in Fig.\ \ref{fig:sincHSRNN-concept}. The sinc filters act as parameterized bandpass filters, initialized with lower and upper cut-off frequencies equally spaced on the Mel scale and windowed with a Hamming window. Since their operation directly corresponds to acoustic bandpass filtering, these filters provide an interpretable and physically realizable front-end that mimics acoustic filtering. The filtered outputs are passed to the HSRNN, which captures temporal dependencies at multiple resolutions.
The sinc filter layer was configured with a kernel size of 101 samples and five output channels. The kernel size defines the temporal window and therefore the frequency resolution of each bandpass filter, while the number of output channels determines how many parallel filters or learnable frequency bands are applied. The filtered outputs are passed to the HSRNN backbone, which consists of three to four recurrent layers with increasing hidden dimensions and a subsampling factor of eight between layers. The final recurrent output is processed by two fully connected layers with tanh activations, followed by a linear output layer and a softmax classifier producing the digit-class probabilities. In the constrained variant, the same structure is used but with non-negative weights, no bias terms, and non-negative activations such as the offset absolute value function. Detailed hyperparameter values and training settings for the SincHSRNNs are listed in Appendix \ref{appendix}.

\subsection{Dataset and preprocessing}
All evaluations were performed on the AudioMNIST dataset \cite{Becker2024}, which contains 30\,000 one-second recordings of spoken digits (0-9) from 60 speakers of different genders and ages. The dataset was divided into a train set (80\%) and a test set (20\%), ensuring that no speaker appears in both sets. This prevents speaker-specific overfitting and provides a realistic benchmark for digit recognition from raw audio.

Prior to training, all audio data underwent several preprocessing steps. Recordings were resampled to lower rates between 1-8\,kHz to reduce sequence length. Each waveform was zero-padded to a fixed duration of one second to ensure consistent input size. Amplitudes were normalized to the range $[-1,1]$, and subsequently squared to represent acoustic intensity. This final transformation aligns the digital representation with the physical interpretation of acoustic intensity, ensuring non-negative inputs consistent with the constraints of acoustic neural networks.

\subsection{Training procedure}
The models were trained using the Adam optimizer \cite{Kingma2015} with cross-entropy loss and a mini-batch size of 64. Adam was selected as it is a widely adopted optimizer that provides stable and efficient convergence in neural network training. Several learning rates were evaluated to ensure stable convergence, leading to the choice of an initial rate of $10^{-3}$ for training and a reduced rate of $10^{-4}$ for fine-tuning during the final epochs. In constrained networks, all weights were clamped to $[0,1]$ after each update, bias terms were omitted, and only non-negative activation functions (offset ReLU or absolute value) were used. Gradient clipping with a maximum norm of 1.0 stabilized training for recurrent models.

A crucial aspect for achieving convergence in the constrained networks was the choice of the weight initialization and activation function offset. Because all weights are restricted to positive values, conventional symmetric initializations (e.g., Xavier or Kaiming) caused many parameters to be clamped to zero, leading to near-constant activations and failed training. To address this, weights were initialized from a non-negative uniform distribution $\mathcal{U}(0,c)$, where the upper bound $c$ was empirically tuned prior to training by analyzing the output activations of an untrained model. The offset parameter in the activation functions was adjusted in conjunction with $c$ to keep activations within a range that avoids both vanishing and saturated regimes.

\subsection{Evaluation metrics}
The model performance was primarily evaluated using classification accuracy on the test set. 
To assess the robustness of the results and estimate statistical uncertainty, each model configuration was trained and evaluated five times with different random initializations. 
Reported accuracies and standard deviations are computed from these five independent runs.

\section{\label{results}Results and discussion}
\subsection{RNNs}
To establish a baseline for speech classification on raw audio data, simple RNNs were trained on the binary subset of AudioMNIST (digits 0 and 1). These models capture temporal dependencies with minimal architectural complexity, making them a suitable test case for assessing the feasibility of acoustic constraints.

Table\ \ref{tab:rnn_results} summarizes the train and test accuracies of unconstrained and constrained RNNs across different hidden-unit configurations at a fixed sampling rate of 1\,kHz. For small networks with 8 or 16 hidden units, both model types achieve comparable test performance of approximately 68\%, with differences well within the run-to-run variability. This shows that the temporal modeling capability of an RNN is largely retained even when constrained to non-negative weights and activations and trained without bias terms. Train and test accuracies remain closely aligned, suggesting minimal overfitting.

As model capacity increases, the constrained networks exhibit a pronounced decline in mean test accuracy and a substantial increase in variability across runs. In particular, architectures with 32 and 64 hidden units show sensitivity to initialization, with some models failing to train and remaining near random-guess performance ($\approx$50\%) throughout all epochs. This instability reflects the limited flexibility introduced by non-negative weight constraints.

\begin{table*}[htbp]
  \begin{ruledtabular}
  \begin{tabular}{lcccc}
    \multirow{2}{*}{\textbf{Hidden units}} &
      \multicolumn{2}{c}{\textbf{Unconstrained}} &
      \multicolumn{2}{c}{\textbf{Constrained}} \\
    & \textbf{Train accuracy (\%)} & \textbf{Test accuracy (\%)} &
      \textbf{Train accuracy (\%)} & \textbf{Test accuracy (\%)} \\
    \colrule
    $8$   & $70.48\pm0.08$ & $68.58\pm0.84$ & $70.10\pm0.51$ & $68.03\pm0.94$ \\
    $16$  & $70.43\pm0.18$ & $68.02\pm0.46$ & $70.17\pm0.37$ & $68.66\pm0.90$ \\
    $32$  & $68.65\pm3.17$ & $68.21\pm1.19$ & $61.35\pm9.36$ & $60.64\pm8.79$ \\
    $64$  & $68.15\pm3.20$ & $66.98\pm2.51$ & $63.31\pm8.17$ & $62.66\pm7.99$ \\
  \end{tabular}
  \end{ruledtabular}
    \caption{\label{tab:rnn_results}Mean train and test accuracies (\%) with standard deviations for unconstrained and physically constrained RNNs on the binary AudioMNIST dataset at a sampling rate of 1\,kHz. While both models achieve comparable performance for smaller architectures, the constrained RNNs display increased variance and reduced robustness as the number of hidden units increases.}
\end{table*}

These findings align with previous work on non-negative neural networks \cite{Becker2023,Hoedt2023,Amos2017}, which reports that such constraints often lead to non-convergent or unstable training unless initialization schemes are adapted to compensate for the restricted parameter space. Because all weights are positive, conventional symmetric initializations result in many parameters being clamped to zero, yielding near-constant activations. This sensitivity underscores the importance of tailored initialization strategies for physically constrained architectures.

With suitable initialization, however, the constrained network achieves performance comparable to the unconstrained model, demonstrating that a physically plausible acoustic RNN can perform binary speech classification directly from raw waveform input. This is encouraging for real-world implementations, where simplicity and physical feasibility are crucial. Nevertheless, it is important to note that these results are obtained on a simplified binary classification task, where the overall accuracy remains well below recent benchmarks for the full AudioMNIST dataset, reporting accuracies up to 98\% using more advanced models and architectures \cite{Becker2024,Tripathi2022,Tukuljac2022}. Conventional RNN approaches typically rely on extracting features such as Mel spectrograms, log-Mel spectrograms, or MFCCs prior to training \cite{Zaman2023}, which reduces input dimensionality and helps mitigate classic RNN challenges like vanishing gradients and limited memory capacity. However, such preprocessing contradicts the design goal of acoustic neural networks, which aim to process raw sound intensities directly.

A natural extension of the RNN is to incorporate mechanisms that improve temporal memory, such as Long Short-Term Memory (LSTM) or Gated Recurrent Unit (GRU) architectures \cite{Hochreiter1997,Cho2014}. However, these models rely on nonlinear activation functions such as sigmoid and $\tanh$, which require finely controlled nonlinearities that are challenging to reproduce in acoustic media. 
Consequently, these architectures are unsuitable for direct physical realization. To overcome the limitations of standard RNNs while maintaining acoustic feasibility, we next investigate HSRNNs \cite{Graves2012}, which provide improved robustness for processing long raw audio sequences.

\subsection{HSRNNs}
To extend the analysis beyond simple recurrent models, HSRNNs \cite{Graves2012} were evaluated on the AudioMNIST dataset at a sampling rate of 1\,kHz, comparing unconstrained and constrained variants across multiple hidden-unit configurations. This architecture introduces temporal downsampling between recurrent layers, enabling efficient modeling of long raw audio sequences while remaining compatible with the acoustic constraints imposed earlier.
\begin{table*}[htbp]
  
  \begin{ruledtabular}
  \begin{tabular}{lcccc}
    \multirow{2}{*}{\textbf{Hidden units}} &
      \multicolumn{2}{c}{\textbf{Unconstrained}} &
      \multicolumn{2}{c}{\textbf{Constrained}} \\
    & \textbf{Train accuracy (\%)} & \textbf{Test accuracy (\%)} &
      \textbf{Train accuracy (\%)} & \textbf{Test accuracy (\%)} \\
    \colrule
    $1$--$1$--$1$       & $61.03\pm9.32$  & $60.76\pm9.03$  & $61.68\pm9.79$  & $60.39\pm8.50$ \\
    $2$--$4$--$8$       & $90.28\pm2.08$  & $89.45\pm1.64$  & $88.04\pm0.97$  & $89.12\pm1.27$ \\
    $4$--$4$--$4$       & $84.47\pm6.43$  & $84.97\pm6.06$  & $71.32\pm1.11$  & $68.99\pm0.70$ \\
    $8$--$8$--$8$       & $91.18\pm1.96$  & $89.71\pm0.70$  & $83.44\pm5.97$  & $82.20\pm8.91$ \\
    $4$--$8$--$16$      & $93.61\pm0.78$  & $90.29\pm0.46$  & $91.42\pm0.56$  & $91.27\pm0.45$ \\
    $8$--$16$--$32$     & $96.91\pm0.58$  & $90.48\pm0.72$  & $92.07\pm1.01$  & $91.30\pm0.79$ \\
    $16$--$32$--$64$    & $99.40\pm0.34$  & $89.85\pm0.63$  & $90.96\pm0.50$  & $89.62\pm1.73$ \\
  \end{tabular}
  \end{ruledtabular}
  \caption{\label{tab:hsrnn_results}Mean train and test accuracies (\%) with standard deviations for HSRNNs on the binary AudioMNIST dataset at a sampling rate of 1\,kHz. Results are reported for unconstrained and physically constrained models across increasing network sizes. Performance generally improves with model capacity.}
\end{table*}

Table \ref{tab:hsrnn_results} summarizes the results for the binary AudioMNIST subset (digits 0 and 1). Configurations are denoted by sequences such as 1-1-1 or 8-16-32, indicating the number of hidden units in the first, second, and third recurrent layers, respectively. Across most configurations, test accuracy remains comparable between the two model types, with the smallest architecture (1-1-1) performing poorly and showing the highest variability across runs. For mid-sized and large configurations, both networks reach accuracies above 85\%, while the constrained model often maintains a narrower gap between train and test accuracy. This reduced discrepancy indicates an implicit regularization effect arising from the non-negativity and absence of bias terms, which limit the model's expressiveness and mitigate overfitting \cite{Sarath2021}. In contrast, the unconstrained networks occasionally reach near-perfect train accuracy, suggesting a tendency toward overfitting.

Extending the evaluation to the full ten-digit AudioMNIST dataset yields the results shown in Tab.\ \ref{tab:hsrnn_results_full}. The overall trends observed in the binary case persist. The smallest constrained configuration (4-8-16) performs significantly worse than the corresponding unconstrained model, with a gap of roughly 18 percentage points, likely due to reduced representational capacity or suboptimal initialization \cite{Becker2023, Kirtas2023, Hoedt2023}. However, for larger networks, the constrained model matches or slightly exceeds the unconstrained model's test accuracy. The best constrained configuration (16-32-64) achieves a mean test accuracy of $66.8\%\pm1.6\%$, representing the highest performance observed under acoustic constraints. Overall, these findings show that hierarchical subsampling improves performance for both constrained and unconstrained networks by enabling efficient modeling of long raw audio sequences.
\begin{table*}[htbp]
  \begin{ruledtabular}
  \begin{tabular}{lcccc}
    \multirow{2}{*}{\textbf{Hidden units}} &
      \multicolumn{2}{c}{\textbf{Unconstrained}} &
      \multicolumn{2}{c}{\textbf{Constrained}} \\
    & \textbf{Train accuracy (\%)} & \textbf{Test accuracy (\%)} &
      \textbf{Train accuracy (\%)} & \textbf{Test accuracy (\%)} \\
    \colrule
    $4$--$8$--$16$         & $66.19\pm0.22$ & $64.62\pm0.42$ & $47.33\pm5.53$ & $47.17\pm6.25$ \\
    $4$--$8$--$16$--$32$   & $72.96\pm1.25$ & $64.45\pm0.57$ & $65.85\pm0.88$ & $63.60\pm1.26$ \\
    $8$--$16$--$32$        & $74.11\pm0.43$ & $66.14\pm0.85$ & $64.58\pm0.69$ & $62.54\pm1.88$ \\
    $16$--$32$--$64$       & $85.61\pm0.27$ & $64.45\pm0.57$ & $71.69\pm0.59$ & $66.79\pm1.57$ \\
  \end{tabular}
  \end{ruledtabular}
  \caption{\label{tab:hsrnn_results_full}Mean train and test accuracies (\%) with standard deviations for HSRNNs evaluated on the full ten-digit AudioMNIST dataset at a sampling rate of 1\,kHz. Results are reported for unconstrained and physically constrained models across increasing network sizes. While constrained models underperform for the smallest configuration, both architectures achieve comparable performance at larger scales, with the constrained model reaching a maximum mean test accuracy of $66.8\%\pm1.6\%$.}
\end{table*}

\begin{figure}[htb]
\centering
\includegraphics[width=\columnwidth]{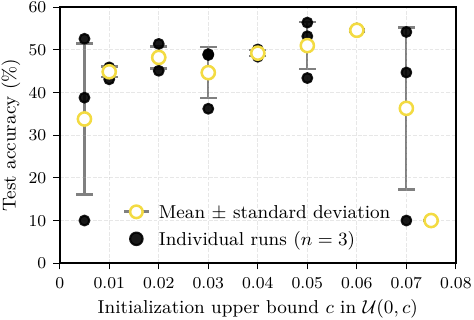}%
\caption{\label{fig:hsrnn-uniform-init} Test accuracy (\%) of the constrained HSRNN as a function of the upper bound $c$ of the uniform weight initialization distribution $\mathcal{U}(0, c)$. Results are shown for the 8-16-32 architecture over three independent runs. Stable training is observed only for intermediate initialization scales ($0.01 \leq c \leq 0.06$), illustrating the high sensitivity of constrained models to weight initialization.}
\end{figure}

Training stability in the constrained networks proved highly sensitive to weight initialization. For the representative 8-16-32 configuration, the corresponding performance shown in Fig.\ \ref{fig:hsrnn-uniform-init} indicates that stable behavior was achieved only for uniform weight initializations $\mathcal{U}(0, c)$ with $c$ between $0.01$ and $0.06$. Larger values for $c$ led to saturation and convergence failure, while smaller values caused vanishing gradients and unstable training. Although the exact range depends on the architecture and activation function, this trend highlights the need for carefully tuned initialization scales when enforcing non-negativity, as inappropriate values can effectively prevent learning.

Training in the constrained HSRNNs led to a pronounced concentration of small weights in the weight matrix. For instance, in the 16-32-64 configuration, approximately 12\% of the weights were clamped to zero after convergence, while the remaining values were strongly biased toward low magnitudes. This distribution indicates a tendency toward sparsity, which arises naturally from the non-negativity constraint. Such behavior is consistent with prior work showing that non-negative constraints promote compact or sparse-like representations \cite{Ayinde2018,Chorowski2015}. Such sparse structures could be advantageous for physical implementations, as they reduce the number of active transmission pathways.

In summary, the HSRNN achieves significantly higher accuracy than the simple RNN baseline while preserving compatibility with physical constraints. The constrained variant benefits from an implicit regularization effect that improves generalization, although it remains sensitive to initialization and exhibits increased variability. These results establish the HSRNN as an effective architecture for processing long raw audio sequences under acoustic feasibility, laying the foundation for the subsequent integration of learnable frequency-selective filters in the SincHSRNN.

\subsection{SincHSRNNs}
To further enhance the classification of raw acoustic signals, we combined the hierarchical subsampling architecture with a learnable sinc-based filter front end \cite{Ravanelli2018}, yielding the SincHSRNN. The sinc filters act as parameterized bandpass filters, providing frequency-selective preprocessing that is both interpretable and physically realizable in acoustic systems. This hybrid design allows the model to learn frequency-domain representations directly from raw waveforms while capturing temporal dependencies at multiple time scales through the hierarchical recurrent structure.

\begin{figure*}[htb]
\centering
\includegraphics[width=1\linewidth]{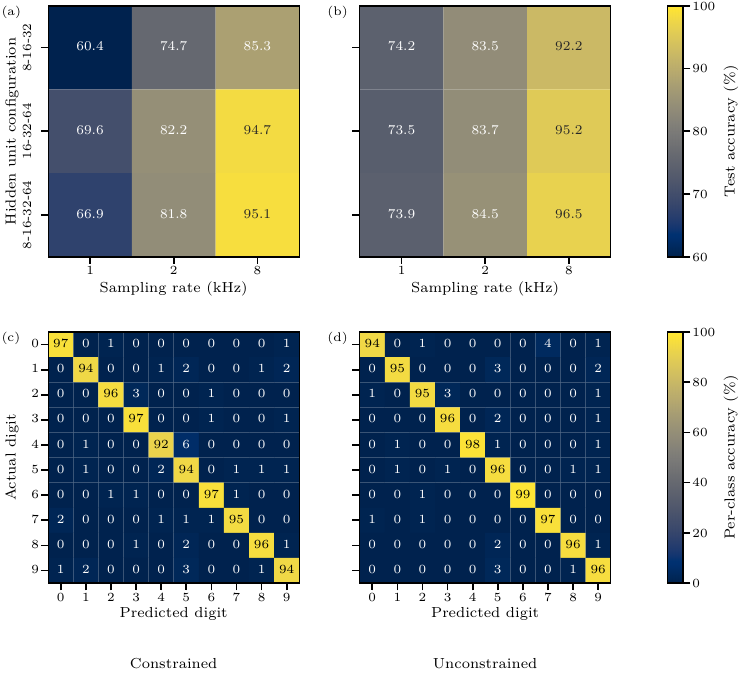}%
\caption{\label{fig:sinchsrnn-heatmap} (a), (b) Test accuracy heatmaps and (c), (d) confusion matrices for (a), (c) constrained and (b), (d) unconstrained SincHSRNNs on the full ten-digit AudioMNIST dataset. (a), (b) Test accuracy heatmaps show the mean test accuracies (\%) across hidden-unit configurations and sampling rates for the constrained and unconstrained networks, respectively. Overall performance increases with sampling rate, and the gap between constrained and unconstrained models narrows at higher capacities. (c), (d) Confusion matrices for the best-performing architecture (8-16-32-64) evaluated at a sampling rate of 8\,kHz, illustrating that both models achieve high per-class accuracy with only minor deviations.}
\end{figure*}

Tables\ \ref{tab:sinchsrnn-unconstrained} and\ \ref{tab:sinchsrnn-constrained}, together with Fig.\ \ref{fig:sinchsrnn-heatmap} summarize the results across different hidden-unit configurations and sampling rates. The test accuracy increases consistently with higher sampling rates for both unconstrained and constrained variants, indicating that finer temporal resolution benefits the learnable filterbank. The confusion matrices in Fig.\ \ref{fig:sinchsrnn-heatmap} further illustrate that the best-performing models achieve uniformly high per-class accuracy, with only minor misclassifications. The highest accuracies are achieved at 8\,kHz with the 8-16-32-64 architecture, reaching $96.5\%$ for the unconstrained and $95.1\%$ for the constrained model.

For the unconstrained networks, the test accuracy varies little with hidden size (differences are below 3\% across architectures at 8\,kHz), suggesting that the performance is primarily limited by input resolution rather than model capacity. The constrained models, in contrast, exhibit stronger dependence on the model size: at 8\,kHz, the accuracy rises from $85.3\%$ to $94.7\%$ when increasing from 8-16-32 to 16-32-64 hidden units. The largest performance gaps between constrained and unconstrained models occur at low sampling rates. At higher sampling rates and larger configurations, however, both models converge to nearly identical accuracies, differing by only a few percentage points.
\begin{table*}[htbp]
  \begin{ruledtabular}
  \begin{tabular}{lccc}
    \textbf{Hidden units} & \textbf{1\,kHz} & \textbf{2\,kHz} & \textbf{8\,kHz} \\
    \colrule
    $8$--$16$--$32$       & $74.15\pm0.43$ & $83.45\pm1.26$ & $92.21\pm1.33$ \\
    $16$--$32$--$64$      & $73.47\pm0.36$ & $83.72\pm0.65$ & $95.23\pm1.81$ \\
    $8$--$16$--$32$--$64$ & $73.94\pm0.45$ & $84.46\pm0.53$ & $96.51\pm1.58$ \\
  \end{tabular}
  \end{ruledtabular}
  \caption{\label{tab:sinchsrnn-unconstrained}Mean test accuracies (\%) with standard deviations for unconstrained sinc-based hierarchical subsampling recurrent neural networks (SincHSRNNs) on the AudioMNIST dataset at sampling rates of 1, 2, and 8\,kHz. Accuracy increases monotonically with sampling rate, indicating that higher temporal resolution enhances the effectiveness of the learned filterbank representation.}
\end{table*}

\begin{table*}[htbp]
  \begin{ruledtabular}
  \begin{tabular}{lccc}
    \textbf{Hidden units} & \textbf{1\,kHz} & \textbf{2\,kHz} & \textbf{8\,kHz} \\
    \colrule
    $8$--$16$--$32$       & $60.35\pm2.55$ & $74.74\pm1.90$ & $85.33\pm1.68$ \\
    $16$--$32$--$64$      & $69.61\pm0.36$ & $82.16\pm1.42$ & $94.70\pm1.39$ \\
    $8$--$16$--$32$--$64$ & $66.86\pm0.82$ & $81.78\pm0.46$ & $95.11\pm1.67$ \\
  \end{tabular}
  \end{ruledtabular}
  \caption{\label{tab:sinchsrnn-constrained}Mean test accuracies (\%) with standard deviations for physically constrained SincHSRNNs on the AudioMNIST dataset at sampling rates of 1, 2, and 8\,kHz. Constrained models exhibit stronger dependence on network size, but approach the performance of unconstrained networks at higher sampling rates, achieving comparable accuracy for the largest configurations.}
\end{table*}

Overall, the SincHSRNN substantially outperforms the pure HSRNN, achieving up to 95\% test accuracy in the constrained setting compared to 67\% for the best HSRNN model. The consistent improvement with the sampling rate indicates that the sinc filters leverage finer temporal resolution to extract frequency-selective features that are inaccessible to the raw recurrent model. The small remaining gap between constrained and unconstrained variants primarily reflects the reduced expressiveness imposed by non-negativity and the absence of bias terms, as discussed previously \cite{Becker2023,Kirtas2023}.

Notably, when evaluated on the full AudioMNIST dataset, the constrained SincHSRNN achieves performance approaching state-of-the-art results from convolutional neural network (CNN)-based architectures \cite{Becker2024,Tripathi2022,Tukuljac2022}, which report accuracies approaching 98\%. As only a limited set of hidden-layer configurations was explored in this study, further optimization of model capacity may lead to additional performance improvements. More recently, transformer-based models have also gained attention in raw audio processing due to their ability to model long-range dependencies \cite{Latif2023}. However, transformer architectures typically involve millions of parameters \cite{Latif2023}, which poses significant challenges for deployment in real-world acoustic systems. In contrast, the SincHSRNN presented in this thesis introduces a novel hierarchical RNN-based approach to raw audio classification that remains compact, using only about 13\% of the parameters of comparable CNNs \cite{Tukuljac2022}. This efficiency and simplicity make it a promising candidate for analog acoustic implementations.

Beyond its accuracy, the architecture offers intrinsic physical interpretability. The learnable sinc filters act as tunable bandpass elements, a standard function in acoustics, allowing a direct mapping between learned filter parameters and physical resonators or waveguide elements. In this sense, the SincHSRNN provides both a high-performing digital model and a plausible design blueprint for real-world acoustic neural networks.

\subsection{Design principles and discussion}

The results presented above reveal several general principles for the design of physically realizable acoustic neural networks. First, the restriction to non-negative weights and activations does not fundamentally limit the ability of the network to learn complex temporal patterns, provided that initialization and scaling are carefully controlled. The reduced overfitting in the constrained HSRNNs and SincHSRNNs demonstrates that non-negativity can act as a form of implicit regularization, promoting sparse representations. This sparsity is advantageous for physical implementations, as it reduces the number of active transmission pathways.

Second, hierarchical subsampling proved essential for effective learning on long raw audio sequences. By aggregating consecutive time steps, the HSRNN progressively reduces the sequence length between recurrent layers. This hierarchical compression alleviates vanishing and exploding gradient effects and lowers the computational load of the recurrent units. In an acoustic realization, the same operation could be achieved by splitting the signal into multiple propagation paths of different lengths and attenuation levels that recombine at a junction. Each path acts as a controllable delay line with a transmission coefficient corresponding to a learnable subsampling weight. 

Third, the integration of learnable sinc filters at the input stage highlights the importance of preprocessing. The sinc-based front end allows the network to learn frequency-selective representations that correspond to acoustic bandpass elements. This not only enhances classification accuracy but also provides a physically meaningful decomposition of the input signal. In an experimental setting, such filters could be realized through arrays of resonators or waveguides tuned to specific frequency bands.

Together, these principles establish a practical framework for developing acoustic neural networks that combine physical feasibility with competitive performance through constrained connectivity, hierarchical temporal integration, and interpretable frequency-selective filtering.

\section{\label{conclusions}Conclusions}
This study establishes a framework for designing and simulating acoustic neural networks, architectures that process information via the propagation of sound waves. By applying physically motivated constraints, including non-negative signals and weights, the omission of bias terms, and nonlinearities compatible with intensity-based, non-negative acoustic signals, we developed digital-twin models that serve as interpretable blueprints for potential analog implementations. Through a systematic progression from recurrent neural networks (RNNs) to hierarchical subsampling RNNs (HSRNNs) and the final SincHSRNN, we demonstrated that networks constrained by physical realizability can achieve high classification accuracy while preserving direct physical interpretability.

The results identify several guiding principles for the development of physically realizable acoustic neural networks. First, non-negativity constraints do not inherently limit representational capacity when initialization and scaling are appropriately controlled. Instead, they promote sparsity and act as an implicit regularization mechanism that enhances generalization. Second, hierarchical subsampling enables robust temporal processing of long sequences and could be physically realized through networks of acoustic delay lines and weighted signal paths. 
Third, the incorporation of learnable sinc filters introduces an interpretable, frequency-selective preprocessing stage that directly corresponds to acoustic bandpass elements. Together, these features allowed the SincHSRNN to reach up to 95\% accuracy on the AudioMNIST dataset, approaching state-of-the-art digital performance while remaining compatible with passive acoustic components.

Beyond their computational performance, these findings highlight the feasibility of acoustic-based analog computation. The learned parameters of the models correspond to measurable material and geometric properties such as attenuation and transmission, establishing a direct link between machine learning representations and physical quantities. This mapping provides a systematic framework for developing low-power, wave-based neural processors that operate through physical signal propagation. In turn, it also enables applications in direct analog speech recognition and other on-device acoustic processing tasks that benefit from low-power, real-time operation.

Future research should focus on translating these digital-twin models into experimental realizations, for instance by constructing small-scale acoustic networks using waveguide structures or resonator arrays that mimic the learned transmission coefficients and nonlinear attenuation effects. While the present work considers a purely passive acoustic system without external energy input, future implementations could in principle include active elements or nonlinear mechanisms capable of signal amplification, allowing effective weights greater than one. However, preliminary simulations indicated that such effects do not provide a meaningful benefit for training stability or performance and were therefore not pursued further here. Extending the simulations to include effects such as scattering, reflection, and noise would refine the mapping between digital parameters and real acoustic responses. On the algorithmic side, developing optimization schemes and initialization strategies specifically tailored for non-negative, physically constrained networks could further enhance stability and convergence. Experimentally, prototype implementations using acoustic metamaterials or other passive wave-based structures could enable direct analog inference on real sound signals, paving the way for low-power edge computing and passive acoustic sensing in environments where electronic systems face limitations.

Overall, this work provides both theoretical foundations and practical design principles for acoustic neural networks, demonstrating that neural computation can, in principle, be achieved through the physics of sound. These results place acoustic systems alongside optical and mechanical analogs as promising candidates for future energy-efficient, wave-guide-based neural computing.

\section*{Supplementary Material}
See Supplementary Material at Ref.\ \citenum{CodeAndData} for the complete source code used to implement and train all neural network models, as well as the raw data and plotting scripts corresponding to the figures presented in this work.

\section*{Data availability}
The data that support the findings of this study are available within the article and its Supplementary Material.

\section*{Conflicts of interest}
There are no conflicts of interest to declare.

\begin{acknowledgments}
R.W.\ and M.R.\ are funded by the Deutsche Forschungsgemeinschaft (DFG, German Research Foundation) -- 535275785 (R.W.); SFB 1459/2 2025 -- 433682494 (R.W., M.R.).
\end{acknowledgments}

\onecolumngrid
\appendix
\section{\label{appendix}}
This appendix provides a comprehensive overview of the training parameters used for all neural network architectures evaluated in this work. 
For each network type, RNN, HSRNN, and SincHSRNN, both the unconstrained and constrained variants were trained using the parameters listed below. 
In constrained models, all weights were clamped to the range $[0,1]$ after each optimization step to emulate acoustic attenuation. The specific hyperparameter values for each architecture are summarized in Tab.\ \ref{tab:params-rnn}--\ref{tab:params-sinchsrn}.

\begin{table*}[htbp]
\begin{ruledtabular}
\begin{tabular}{lcc}
\textbf{Parameter} & \textbf{Unconstrained network} & \textbf{Constrained network} \\
\hline
Epochs & 30 & 30 \\
Batch size & 64 & 64 \\
Learning rate & 0.001 & 0.001 \\
Loss function & Cross-entropy loss & Cross-entropy loss \\
Optimizer & Adam & Adam \\
Bias & True & False \\
Weight initialization & $\mathcal{U}_\text{Xavier}(-a,\,a)$  & $\mathcal{U}(0, c),~c\in[0.01,\,0.1]$\\
Activation function & $\tanh$ & $|x - c|$ or $\max(x - c, 0), ~c\in[0.01,\,0.1]$ \\
\end{tabular}
\end{ruledtabular}
\caption{\label{tab:params-rnn}Training parameters for the RNNs used in the AudioMNIST binary classification task. The unconstrained network follows a standard Xavier uniform initialization $\mathcal{U}_\text{Xavier}(-a, a)$ and employs a $\tanh$ activation. The constrained network uses non-negative weights sampled from $\mathcal{U}(0, c)$ with $c\in[0.01,\,0.1]$  and replaces the activation with offset absolute or offset ReLU functions $|x-c|$ or $\max(x-c,0)$, respectively.}
\end{table*} 

\begin{table}[htbp]
\begin{ruledtabular}
\begin{tabular}{lcc}
\textbf{Parameter} & \textbf{Unconstrained network} & \textbf{Constrained network} \\
\hline
Epochs & 50 & 50-200 \\
Batch size & 64 & 64 \\
Learning rate & 0.001 & 0.001 \\
Hidden layers & 3 & 3\\
Loss function & Cross-entropy loss & Cross-entropy loss \\
Optimizer & Adam & Adam \\
Bias & True & False \\
Weight initialization & $\mathcal{U}_\text{Xavier}(-a,\,a)$  & $\mathcal{U}(0, c),~c\in[0.05,\,0.3]$ \\
Activation function & $\tanh$ & $|x - c|$ or $\max(x - c, 0), ~c\in[0.05,\,0.3]$\\
\end{tabular}
\end{ruledtabular}
\caption{\label{tab:params-hsrnn}Training parameters for the HSRNNs used in the AudioMNIST classification task. The unconstrained network employs standard Xavier uniform initialization $\mathcal{U}_\text{Xavier}(-a, a)$ with $\tanh$ activations.  The constrained variant enforces non-negative weights drawn from $\mathcal{U}(0, c)$ with $c\in[0.05,\,0.3]$, replaces the activation by offset absolute or offset ReLU functions $|x-c|$ or $\max(x-c,0)$, and removes bias terms. Longer training schedules (50-200 epochs) were required for convergence under these constraints.}
\end{table}

\begin{table}[htbp]
\begin{ruledtabular}
\begin{tabular}{lcc}
\textbf{Parameter} & \textbf{Unconstrained network} & \textbf{Constrained network} \\
\hline
Epochs & 40 & 60-95 \\
Batch size & 64 & 64 \\
Learning rate & 0.001, 0.0001 & 0.001, 0.0001 \\
Hidden layers & 3-4 & 3-4\\
Loss function & Cross-entropy loss & Cross-entropy loss \\
Optimizer & Adam & Adam \\
Bias & True & False \\
Subsampling factor & 8 & 8\\
Sinc kernel size& 101 & 101\\
Sinc output channels& 5 & 5\\
Weight initialization & $\mathcal{U}_\text{Xavier}(-a,\,a)$ & $|\mathcal{U}_\text{Xavier}(-a,\,a)|$, gain $=0.13$ \\
Activation function & $\tanh$ & $|x - c|, ~c\in[0.9,\,1]$\\
\end{tabular}
\end{ruledtabular}
\caption{\label{tab:params-sinchsrn}Training parameters for the SincHSRNN used in the AudioMNIST classification task. 
The unconstrained network follows standard Xavier uniform initialization $\mathcal{U}_\text{Xavier}(-a, a)$ with $\tanh$ activations. The constrained variant employs non-negative weights obtained from the absolute Xavier uniform distribution $|\mathcal{U}_\text{Xavier}(-a, a)|$ with a gain factor of 0.13, and uses offset absolute activations $|x-c|$ with $c\in[0.9,\,1]$. Bias terms were omitted. Training was performed for 60-95 epochs, with the final 10 epochs at a reduced learning rate 0.0001 for fine-tuning.}
\end{table}

\FloatBarrier
\twocolumngrid
\nocite{apsrev41Control}
\bibliographystyle{apsrev4-1}
\bibliography{control,refs}

\end{document}